\documentclass[a4paper,superscriptaddress,twocolumn,showpacs,prl]{revtex4}%
\usepackage{mathrsfs}
\usepackage{amsfonts}
\usepackage{amsmath}
\usepackage{amssymb}
\usepackage{graphicx}
\usepackage{float}
\usepackage{braket,hyperref}%
\providecommand{\U}[1]{\protect\rule{.1in}{.1in}}
\begin{document}
\title{Entangled Terahertz Photon Emission from Topological Insulator Quantum Dots}
\author{Li-kun Shi}
\affiliation{Beijing Computational Science Research Center, Beijing 100193, China}
\author{Kai Chang}
\affiliation{SKLSM, Institute of Semiconductors, Chinese Academy of Sciences, P.O. Box 912,
Beijing 100083, China}
\author{Chang-Pu Sun}
\affiliation{Beijing Computational Science Research Center, Beijing 100193, China}

\begin{abstract}
We propose an experimentally feasible scheme for generating entangled
terahertz photon pairs in topological insulator quantum dots (TIQDs). We
demonstrate theoretically that in generic TIQDs: 1) the fine structure
splitting, which is the obstacle to produce entangled photons in conventional
semiconductor quantum dots, is inherently absent for one-dimensional massless
excitons due to the time-reversal symmetry; 2) the selection rules obey
winding number conservation instead of the conventional angular momentum
conservation between edge states with a linear dispersion. With these two
advantages, the entanglement of the emitted photons during the cascade in our
scheme is robust against unavoidable disorders and irregular shapes of the TIQDs.

\end{abstract}

\pacs{03.67.Bg, 71.70.Gm, 73.21.La, 78.67.-n}
\maketitle

\textit{Introduction.---} Entanglement is the heart of quantum physics and
quantum information technology\cite{BB84,Ekert,Bennett}. Semiconductor quantum
dots (QDs), named as artificial atoms, are one of the most promising sources
for quantum light, e.g., single photon and entangled photon pair
\cite{Benson,Stevenson-Nature,Young,Akopian,Hudson,Singh,Schliwa,Salter,Recher,AJBennett,Mohan,Kuklewicz,Trotta1,Juska,Muller,Versteegh,Trotta2,JPWang}%
, because, comparing with real atoms and other sources, QDs possess an unique
advantage of being compatible with well-developed semiconductor
integration\cite{Juska,Versteegh} as well as on-demand and non-probabilistic
feature\cite{Stevenson-Nature,Young,Muller} of biexciton cascade emission in
QDs. The fine structure splitting (FSS), the energy difference existing
between two bright exciton states in QDs with unavoidable irregular shapes,
strongly spoils the entanglement of the emitted photons\cite{Hudson}. To
overcome the FSS brought by breaking of rotational symmetry, various methods
such as sophisticated QD growth
control\cite{Singh,Schliwa,Mohan,Juska,Versteegh}, external
electric\cite{AJBennett}, magnetic\cite{Hudson}, and strain
fields\cite{Kuklewicz,Trotta1,Trotta2,JPWang}, have been used to manipulate
and/or suppress the FSS of excitons in semiconductor QDs. However, for
practical devices with large numbers of QDs differ from dot to dot, it is a
great challenge to scale all the QDs at the same time. Therefore, QDs having
no FSS intrinsically, regardless of random alloy distributions and other
uncontrollable effects, are essential to realize integrated entangled photon emitters.

In this Letter, we show emitters composed of topological insulator quantum
dots (TIQDs) can produce terahertz photon pairs with robust entanglement. We
first present the theory in detail with a HgTe QD to illustrate our main ideas
and notations. Then we prove by the time-reversal symmetry (TRS) and winding
number conservation, one-dimensional (1D) massless excitons in generic TIQDs
with random disorders and morphology fluctuations are free from FSS, and the
entanglement of the produced photon pairs from the cascade process is robust.

\begin{figure}[tbh]
\includegraphics[width=0.95\columnwidth]{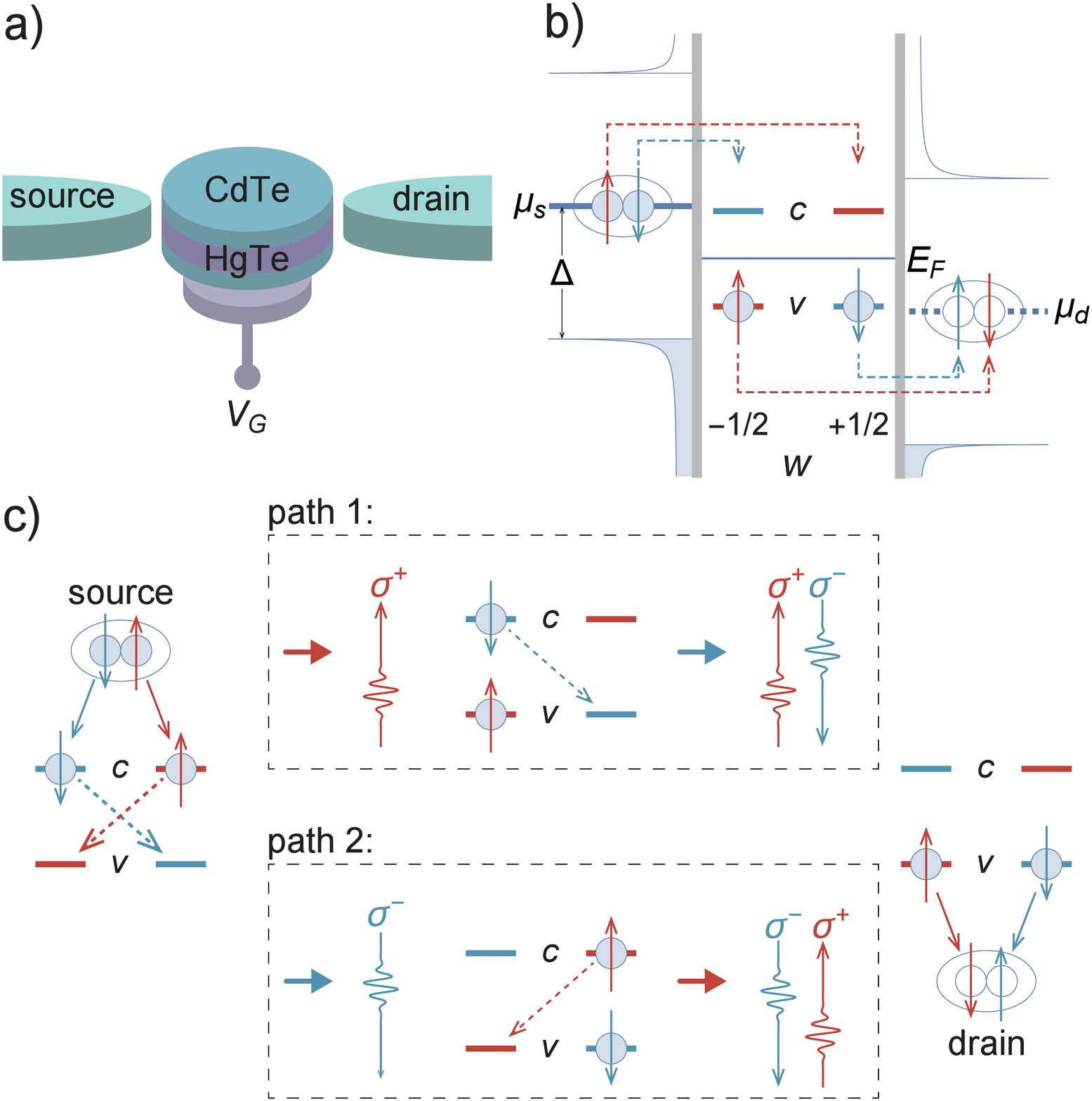}\caption{(color online). (a)
Sketch of the proposed $p$-$n$ junction with SC source and drain electrodes
weakly coupled to a HgTe QD whose Fermi level is controlled by a gate
electrode $V_{G}$. (b) Energy diagram for the structure shown in (a): $\Delta$
is the SC gap; $\mu_{\mathrm{s}}$ and $\mu_{\mathrm{d}}$ are the Fermi
energies of the source and drain electrodes; $E_{F}$ is the Fermi level and
$c$ ($v$) stands for conduction (valance) level we defined in the HgTe QD; Red
(blue) bars inside the QD region are spin-up (down) edge states, $w$ denotes
their winding numbers; Shaded circles with red (blue) arrows are spin-up
(down) electrons; The white circles on the drain side are holes; The ellipses
enclose two shaded (white) circles are electron- (hole-) type Cooper pairs;
Dashed arrows show the charge transfers from the source to drain electrode
through the HgTe QD as electron singlets. The thin grey areas are barriers
between the leads and the QD. (c) The initial state of the QD $\left\vert
2\right\rangle _{c}\left\vert 0\right\rangle _{v}$ goes through two
indistinguishable pathes with the intermediate state $\left\vert \sigma
^{+}\right\rangle \left\vert \downarrow\right\rangle _{c}\left\vert
\uparrow\right\rangle _{v}+\left\vert \sigma^{-}\right\rangle \left\vert
\uparrow\right\rangle _{c}\left\vert \downarrow\right\rangle _{v}$, and
becomes an entangled final state $\left(  \left\vert \sigma^{+}\sigma
^{-}\right\rangle +\left\vert \sigma^{-}\sigma^{+}\right\rangle \right)
\left\vert 0\right\rangle _{c}\left\vert 2\right\rangle _{v}/\sqrt{2}$.
$\sigma^{\pm} $ is a right-handed (left-handed) circularly polarized photon
carrying one unit of angular momentum $\pm\hbar$. Dashed arrows are possible
decay pathes.}%
\label{fig1}%
\end{figure}

\textit{Topological insulator quantum dots.---}We consider a
superconductor-TIQD hybrid system\cite{Franceschi} as an example: a TIQD is
fabricated in a CdTe/HgTe/CdTe heterostructure by etching technique, and it is
weakly coupled to superconducting (SC) leads on two sides at $p$- or $n$-type
regime, respectively [see Fig. \ref{fig1} (a)]. The low-energy electron states in the \textit{isolated} HgTe QD
are described by a four-band effective Hamiltonian in the basis $\left\vert
e,\uparrow\right\rangle $, $\left\vert hh,\uparrow\right\rangle $, $\left\vert
e,\downarrow\right\rangle $, and $\left\vert hh,\downarrow\right\rangle
$\cite{Bernevig,hh}:%
\begin{equation}
H_{4\times4}=\left[
\begin{array}
[c]{cc}%
H_{\uparrow}(\hat{\mathbf{k}}) & 0\\
0 & H_{\downarrow}(\hat{\mathbf{k}})
\end{array}
\right]  +V\left(  r\right)  ,
\end{equation}
where $\hat{\mathbf{k}}=(\hat{k}_{x},\hat{k}_{y})$ is the in-plane wavevector,
$H_{\uparrow}(\hat{\mathbf{k}})=H_{\downarrow}^{\ast}(-\hat{\mathbf{k}%
})=\epsilon(\hat{k})I_{2\times2}+d_{i}(\hat{\mathbf{k}})\sigma^{i}$, and
$\sigma^{i}\left(  i=1,2,3\right)  $ are the Pauli matrices; $\epsilon(\hat
{k})=C-D\hat{k}^{2}$, $d_{1,2}(\hat{\mathbf{k}})=A\hat{k}_{x,y} $, and
$d_{3}(\hat{k})=M-B\hat{k}^{2}$. The parameters $A$, $B$, $C$, $D$, and $M$
depend on the thickness of the HgTe layer. $M$ is used to describe the band
insulator with a positive gap $\left(  M>0\right)  $ and non-trivial
topological insulator with a negative gap $\left(  M<0\right)  $. The
hard-wall confining potential $V\left(  r\right)  $ is employed to describe
potential difference between the vacuum and the TIQD with $R$ as its radius.
The eigenstates of the QD $\left\vert \psi_{j,\uparrow\downarrow}\right\rangle
$\cite{Tsitsishvili,SM} are linear combinations of degenerate eigenstates of
$\hat{J}_{z}$ which is conserved along the $z$ axis. Note, we refer to
$\left\vert \psi_{j,\uparrow\downarrow}\right\rangle $ as spin-up (down)
states, as they are the eigenstates for the upper (lower) block of the
$4\times4$ Hamiltonian.

For a given $w_{\uparrow\downarrow}=j\mp1$, the lowest energy state
$\left\vert w,\uparrow\downarrow\right\rangle =\left\vert \psi_{w\pm
1,\uparrow\downarrow}\right\rangle _{\mathrm{low}}$ with its energy
$\epsilon_{w,\uparrow\downarrow}$ in the non-trivial gap $[M,-M]$, is a
massless edge state\cite{KChang,JLi}. Here we go beyond Refs.
\cite{KChang,JLi} and use winding numbers to classify the edge states:
$2w_{\uparrow\downarrow}=(j\mp1/2)+(j\mp3/2)$ is the total winding number in
the basis $\left\vert e,\uparrow\downarrow\right\rangle $ and $\left\vert
hh,\uparrow\downarrow\right\rangle $ as we will explain below. The edge states
localize on the edge and possess a linear energy dispersion $\epsilon
_{w,\uparrow\downarrow}=\pm w\cdot\Delta\epsilon+\epsilon_{0}$. The energy
level spacing $\Delta\epsilon=-A[1-\left(  D/B\right)  ^{2}]^{1/2}R^{-1}%
$\cite{KChang} depends on QD's radius, and $\epsilon_{0}$ is the constant sets
the Dirac point. $\left\vert w,\uparrow\right\rangle $ and $\left\vert
-w,\downarrow\right\rangle =\Theta\left\vert w,\uparrow\right\rangle $ are
always degenerate due to the TRS. Here $\Theta=-i\sigma_{y}K$ is a
time-reversal operator and $K$ is the operation of complex conjugation. Due to
the TRS, the backscattering between $\left\vert w,\uparrow\right\rangle $ and
$\left\vert -w,\downarrow\right\rangle $ are not allowed, therefore edge
states possess a long spin lifetime\cite{Hasan,XLQi}. The momentum matrix
$\mathbf{P}\propto\partial H_{4\times4}/\partial\mathbf{k}$ is block-diagonal
and will not couple the spin-up and spin-down states, thus the transition
between the edge states with opposite spins is forbidden. For photons
propagating in $z$ direction, circular polarized transitions only exist
between the states with the same spin while conserving $j$.

We assume the Fermi energy of the HgTe QD is between two energy levels
$\epsilon_{c}=\epsilon_{\pm1/2,\uparrow\downarrow}$ and $\epsilon_{v}%
=\epsilon_{\mp1/2,\uparrow\downarrow}$. They are evenly away from the Dirac
point, and the corresponding edge states have an additional TRS symmetry
$i\Theta\left\vert w,\uparrow\downarrow\right\rangle =$ $\left\vert
-w,\uparrow\downarrow\right\rangle $ because near the Dirac point keeping the
linear terms in $k$ one has $K[\sigma_{y}H_{\uparrow\downarrow}(\hat
{\mathbf{k}})\sigma_{y}\cdot\sigma_{y}\left\vert w,\uparrow\downarrow
\right\rangle ]=-H_{\uparrow\downarrow}(\hat{\mathbf{k}})[i\Theta\left\vert
w,\uparrow\downarrow\right\rangle ]=\epsilon_{w,\uparrow\downarrow}%
[i\Theta\left\vert w,\uparrow\downarrow\right\rangle ]$. We call $\epsilon
_{c}$ conduction level and $\epsilon_{v}$ valance level for convenience.
$H_{\alpha}=\epsilon_{\alpha}\sum_{\sigma=\uparrow\downarrow}a_{\alpha,\sigma
}^{\dag}a_{\alpha,\sigma}+U_{\alpha}n_{\alpha,\uparrow}n_{\alpha,\downarrow}$
is the Hamiltonian for the conduction ($\alpha=c$) or valance ($\alpha=v$)
level. Two pairs of degenerate edge states are obtained by the creation or
annihilation operator: $a_{\alpha,\uparrow\downarrow}^{\dag}\left\vert
0\right\rangle _{c}=$ $a_{\alpha,\uparrow\downarrow}\left\vert 2\right\rangle
_{\alpha}=\left\vert \downarrow\uparrow\right\rangle _{\alpha}$, where
$\left\vert 0\right\rangle _{\alpha}$ is the empty state and $\left\vert
2\right\rangle _{\alpha}=a_{\alpha,\uparrow}^{\dag}a_{\alpha,\downarrow}%
^{\dag}\left\vert 0\right\rangle _{\alpha}$ is the two-electron occupied state
for the conduction/valance ($\alpha=c/v$) level. We denote $\left\vert
\uparrow\downarrow\right\rangle _{c}=\left\vert \pm1/2,\uparrow\downarrow
\right\rangle $ and $\left\vert \uparrow\downarrow\right\rangle _{v}%
=\left\vert \mp1/2,\uparrow\downarrow\right\rangle $ since the spin and
angular momentum indices are locked for the corresponding level. We will prove
later: 1) when there is one electron in the conduction level and one electron
in the valance level, the resulting two-particle states $\left\vert
\uparrow\right\rangle _{c}\left\vert \downarrow\right\rangle _{v}$ and
$\left\vert \downarrow\right\rangle _{c}\left\vert \uparrow\right\rangle _{v}$
are degenerate exciton states with the same Coulomb repulsion $U$ between the
two electrons due to the TRS; 2) $U_{c,v}=U$ describes the charging energy by
electron edge states in the conduction/valance level.

\textit{Superconducting} $p$\textit{-}$n$ \textit{junction as a light-emitting
diode.---} We coupled the HgTe QD weakly to a $s$-wave SC $p$-$n$ junction,
with pair potential $\Delta_{\mathrm{s},\mathrm{d}}$ and chemical potential
$\mu_{\mathrm{s},\mathrm{d}}$ for its source (drain) lead [see Fig. \ref{fig1}
(b)]. $\mathrm{e}V_{\mathrm{sd}}=\mu_{\mathrm{s}}-\mu_{\mathrm{d}}$ is the
bias voltage applied across the junction. $\mu_{\mathrm{s},\mathrm{d}}$ is
aligned to the conduction (valance) level $\epsilon_{c,v}$. Once the charge
transfers through the $p$-$n$ junction, biexciton state $\left\vert
2\right\rangle _{c}\left\vert 0\right\rangle _{v}$, which has two electrons in
the conduction level and no electron in the valance level, would go through a
two-step cascade process [see Fig. \ref{fig1} (c)]: 1) $\left\vert
2\right\rangle _{c}\left\vert 0\right\rangle _{v}\rightarrow\left\vert
1\right\rangle _{c}\left\vert 1\right\rangle _{v}$, the biexciton undergoes a
spontaneous\ transition to $\left\vert 1\right\rangle _{c}\left\vert
1\right\rangle _{v}$, one of the two degenerate exciton states $\left\vert
\uparrow\right\rangle _{c}\left\vert \downarrow\right\rangle _{v}$ and
$\left\vert \downarrow\right\rangle _{c}\left\vert \uparrow\right\rangle _{v}%
$; 2) $\left\vert 1\right\rangle _{c}\left\vert 1\right\rangle _{v}%
\rightarrow\left\vert 0\right\rangle _{c}\left\vert 2\right\rangle _{v}$, the
exciton state $\left\vert 1\right\rangle _{c}\left\vert 1\right\rangle _{v}$
decays to $\left\vert 0\right\rangle _{c}\left\vert 2\right\rangle _{v}$.
During the cascade, two photons with energy $\hbar\omega_{0}\simeq
\mathrm{e}V_{\mathrm{sd}}$, which is in terahertz ($\sim\mathrm{meV)}$, are
produced. Here we use a waveguide to produce photons propagating along the $z$
axis only\cite{Versteegh,KWang}. The transition is described by the following
Hamiltonian:%
\begin{equation}
H_{\mathrm{int}}=g\sum_{q_{z}}\left(  \hat{b}_{q_{z},+}^{\dag}a_{v,\uparrow
}^{\dag}a_{c,\uparrow}+\hat{b}_{q_{z},-}^{\dag}a_{v,\downarrow}^{\dag
}a_{c,\downarrow}\right)  +\mathrm{h.c.},
\end{equation}
which conserves the $J_{z}$, as a right-handed (left-handed) circularly
polarized photon $\hat{b}_{q_{z},\pm}^{\dag}\left\vert 0\right\rangle
=\left\vert \sigma^{\pm}\right\rangle $ conveys one unit of angular momentum
$\pm\hbar$. The coupling strength $g\sim e(\hbar\omega_{0}/2\epsilon V)^{1/2}$
can be tuned by changing the volume of the waveguide. After each biexciton
cascade, the final two-photon state entangled with opposite circular
polarizations is%
\begin{equation}
\left\vert \psi\right\rangle =\left(  \left\vert \sigma^{+}\sigma
^{-}\right\rangle +\left\vert \sigma^{-}\sigma^{+}\right\rangle \right)
/\sqrt{2}.
\end{equation}
This final state means the two sequentially produced photons in the $z$
direction are polarization entangled: if the first emitted photon is
$\sigma^{+}$ ($\sigma^{-}$), then the second emitted photon is $\sigma^{-}$
($\sigma^{+}$). For a HgTe QD with a radius $R=150$\textrm{nm}, both the
biased voltage $\mathrm{e}V_{\mathrm{sd}}\simeq2.4\mathrm{meV}$ and the
charging energy $U\simeq0.8\mathrm{meV}$\cite{JLi} are smaller than
$\Delta_{\mathrm{s},\mathrm{d}}\approx3\mathrm{meV}$ (for Nb). In this case,
we neglect the transitions creating a single quasiparticle in the SC leads.
The emission intensity of circular polarization entangled photons equals to
charge current through the junction.

The weak coupling between the HgTe QD and the SC leads will slightly split the
energy levels for two-particle states $\left\vert 2\right\rangle
_{c}\left\vert 0\right\rangle _{v}$ and $\left\vert 0\right\rangle
_{c}\left\vert 2\right\rangle _{v}$ and shifts the frequencies of emitted
photons from the central frequency $\omega_{0}$\cite{Recher,SM}. However, edge
states and exciton states in the HgTe QD is unaffected by this weak coupling,
and the polarizations and energies of the photons are uncorrelated. The delay
time $\tau_{X}$ between the first and second emission events is much shorter
than the emission time of the pair $\tau_{XX}$ ($\tau_{X}/\tau_{XX}\sim
0.09$)\cite{Recher}, thus the two photons in a pair are time- and
polarization-correlated. In our proposal, the SC gaps forbid single
quasiparticle creation in the SC leads, so the entangled photons are produced
pair by pair and could be identified efficiently in the time domain. This
pair-by-pair feature could also appear in a large non-trivial gap (about
$0.3$\textrm{eV}) TIQD\cite{Yong} with the aid of the Pauli exclusion
principle\cite{Benson}.

Superradiance and Josephson effect has been discussed in similar SC $p$-$n$
systems\cite{Recher,Asano}. Here we neglect these effects because the absence
of a static in-plane electric field does not allow the emission of a single
photon from Cooper pair transfer through the QD via Josephson
effect\cite{Recher}. Our proposal is based on the biexciton cascade process
which is not related to the Josephson effect.\begin{figure}[tbh]
\includegraphics[width=0.86\columnwidth]{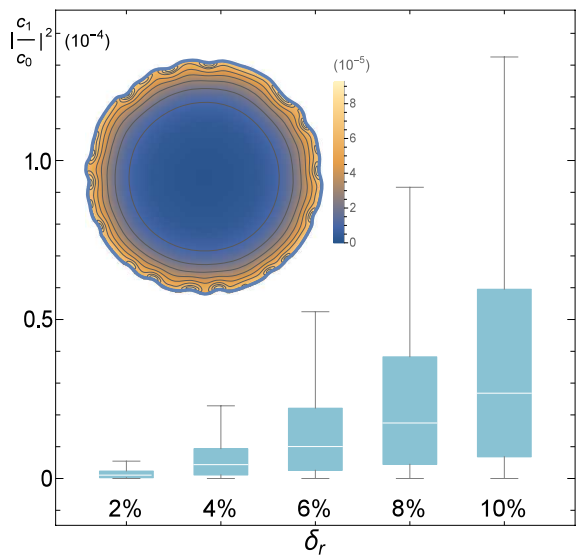}\caption{(color online).
Calculated $|c_{1}/c_{0}|^{2}$ for $10^{3}$ TIQDs using finite element method.
The TIQDs are generated by fluctuating their radius randomly as $r_{0}\left[
1+\delta_{r}f\left(  \theta\right)  \right]  $ in the polar coordinate system;
$r_{0}=150$nm, $f\left(  \theta\right)  \in\left[  -1,1\right]  $ is a random
function. The results are classified by $\delta_{r}$ which denotes the maximum
fluctuation percentage. The inset shows the density distributions of an edge
state in one of the randomly shaped TIQDs.}%
\label{fig2}%
\end{figure}

\textit{Zero FSS in generic TIQDs.---} We now prove in generic TIQDs,
two-particle states $\left\vert \uparrow\right\rangle _{c}\left\vert
\downarrow\right\rangle _{v}$ and $\left\vert \downarrow\right\rangle
_{c}\left\vert \uparrow\right\rangle _{v}$ are degenerate exciton states
without FSS. $\left\vert \uparrow\right\rangle _{c}\left\vert \downarrow
\right\rangle _{v}$ and $\left\vert \downarrow\right\rangle _{c}\left\vert
\uparrow\right\rangle _{v}$ are products of edge states which satisfy: 1)
$\left\vert \downarrow\right\rangle _{\alpha}=\left\vert \Theta\uparrow
\right\rangle _{\alpha}$ ($\alpha=c,v$) due to the TRS; 2) more specifically,
an additional TRS symmetry $\left\vert \uparrow\downarrow\right\rangle
_{c}=i\Theta\left\vert \uparrow\downarrow\right\rangle _{v}$ near the Dirac
point. These two relations hold, not only for HgTe QDs, but also for generic
TIQDs that are composed of other materials.

The two-particle Hamiltonian for the conduction and valance level is
$H_{cv}=\sum_{\alpha=c,v}\sum_{\sigma=\uparrow\downarrow}\epsilon_{\alpha
}a_{\alpha,\sigma}^{\dag}a_{\alpha,\sigma}+\hat{V}$, where $\hat{V}$ is the
Coulomb interaction between the two electrons. The bright exciton states must
be the linear combinations of the two-particle states $\left\vert
\uparrow\right\rangle _{c}\left\vert \downarrow\right\rangle _{v}$ and
$\left\vert \downarrow\right\rangle _{c}\left\vert \uparrow\right\rangle _{v}%
$\cite{SM}. In the basis $\{\left\vert \uparrow\right\rangle _{c}\left\vert
\downarrow\right\rangle _{v},\left\vert \downarrow\right\rangle _{c}\left\vert
\uparrow\right\rangle _{v}\}$, the off-diagonal term of $H_{cv}$ is
$\left\langle \uparrow\right\vert _{c}\left\langle \downarrow\right\vert
_{v}H_{cv}\left\vert \downarrow\right\rangle _{c}\left\vert \uparrow
\right\rangle _{v}=\left\langle \uparrow\right\vert _{c}\left\langle
\downarrow\right\vert _{v}\hat{V}\left\vert \downarrow\right\rangle
_{c}\left\vert \uparrow\right\rangle _{v}=0$\cite{SM} as $\hat{V}$ commutes
with the time-reversal operator $\Theta^{2}=-1$. Two diagonal terms are also
the same as $\left\langle \uparrow\right\vert _{c}\left\langle \downarrow
\right\vert _{v}\hat{V}\left\vert \uparrow\right\rangle _{c}\left\vert
\downarrow\right\rangle _{v}=\left\langle \downarrow\right\vert _{c}%
\left\langle \uparrow\right\vert _{v}\hat{V}\left\vert \downarrow\right\rangle
_{c}\left\vert \uparrow\right\rangle _{v}\equiv U$. Therefore $\left\vert
\uparrow\right\rangle _{c}\left\vert \downarrow\right\rangle _{v}$ and
$\left\vert \downarrow\right\rangle _{c}\left\vert \uparrow\right\rangle _{v}$
are degenerate exciton states with the same energy $\epsilon_{c}+\epsilon
_{v}+U$. Besides, $U_{\alpha}=\left\langle \uparrow\right\vert _{\alpha
}\left\langle \downarrow\right\vert _{\alpha}\hat{V}\left\vert \uparrow
\right\rangle _{\alpha}\left\vert \downarrow\right\rangle _{\alpha}=U$
($\alpha=c,v$)\cite{SM}.

On the other side, the exciton FSS arises from the exchange interaction
between two exciton states. Note that, by swapping its electron with the hole,
$\left\vert \uparrow\right\rangle _{c}\left\vert \downarrow\right\rangle _{v}$
is transformed to $\left\vert \downarrow\right\rangle _{c}\left\vert
\uparrow\right\rangle _{v}$. The exchange energy\cite{Bayer} $E_{\mathrm{ex}%
}\propto\left\langle \uparrow\right\vert _{c}\left\langle \downarrow
\right\vert _{v}\hat{V}\left\vert \downarrow\right\rangle _{c}\left\vert
\uparrow\right\rangle _{v}=0$ leads to vanishing FSS between the two exciton
states. The physical reason behind the zero FSS is that the backscattering is
forbidden by TRS. It is necessary to emphasize that, the unavoidable
non-magnetic disorders and morphology variations of a TIQD will not break TRS.
Therefore the FSS vanishes between exciton states $\left\vert \uparrow
\right\rangle _{c}\left\vert \downarrow\right\rangle _{v}$ and $\left\vert
\downarrow\right\rangle _{c}\left\vert \uparrow\right\rangle _{v}$ in generic
TIQDs as long as the TRS symmetry exist. This is the significance between the
TIQDs and the conventional semiconductor QDs studied before.

\textit{Winding number conservation.--- }One may intuitively conclude
$SO\left(  2\right)  $ symmetry in an actual TIQD will be broken, so that
$J_{z} $ is not conserved and the original selection rules do not hold.
Therefore, the final photon state $\left\vert \psi^{\prime}\right\rangle
=c_{0}\left(  \left\vert \sigma^{+}\sigma^{-}\right\rangle +\left\vert
\sigma^{-}\sigma^{+}\right\rangle \right)  /\sqrt{2}+c_{1}\left(  \left\vert
\sigma^{+}\sigma^{+}\right\rangle +\left\vert \sigma^{-}\sigma^{-}%
\right\rangle \right)  /\sqrt{2}$ with $\left\vert c_{0}\right\vert
^{2}+\left\vert c_{1}\right\vert ^{2}=1$ possesses the reduced fidelity
$\mathcal{F}=[1+|c_{1}/c_{0}|^{2}]^{-1}$ with respect to Bell state. To
quantitatively analyze the magnitude of $|c_{1}/c_{0}|^{2}$, we have performed
finite element method simulations\cite{SM} on about $10^{3}$ different HgTe
QDs with random shapes. We generate these QDs by fluctuating their boundaries
as $r\left(  \theta\right)  =r_{0}[1+\delta_{r}f\left(  \theta\right)  ]$ in
the polar coordinate system, where the mean radius is $r_{0}=\int_{0}^{2\pi
}r\left(  \theta\right)  \mathrm{d}\theta/2\pi$ while the random function
$f\left(  \theta\right)  \in\left[  -1,1\right]  $ satisfies $f\left(
\theta\right)  =f\left(  \theta+2\pi\right)  $, where $\delta_{r}$ is a
dimensionless number characterizing the maximum fluctuation percentage
(typically less than $10\%$ in experiments). The simulation results (see Fig.
\ref{fig2}) show that the irregularities of the TIQDs' configurations only
cause neglectable deviations ($\lesssim10^{-4}$) of selection rules from that
with a perfect circular shape.

\begin{figure}[tbh]
\includegraphics[width=0.6\columnwidth]{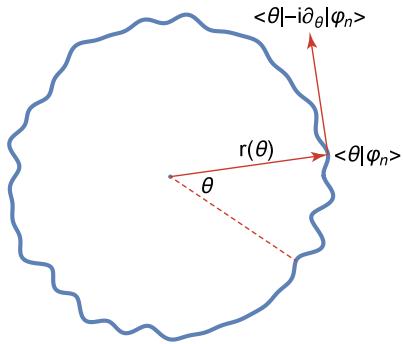}\caption{(color online). A 1D
system on a closed loop with linear dispersion. $\left\langle \varphi
_{n}\right\vert \hat{k}_{\theta}\left\vert \varphi_{n}\right\rangle $ is the
winding number counting the total number of counterclockwise turns that
$\left\vert \varphi_{n}\right\rangle $ makes around the origin.}%
\label{fig3}%
\end{figure}

We have a topological explanation for this robustness of selection rules
against randomness of the TIQDs' morphologies. Consider a 1D system on a
closed loop with \textit{linear} dispersion (see Fig. \ref{fig3}).
The Hamiltonian for the system reads $\hat{H}=\Delta\epsilon
\cdot\hat{k}_{\theta}+V\left(  \theta\right)  $ and depends on a single
parameter $\theta\in\left[  0,2\pi\right]  $. Both the variations of the
radius $r\left(  \theta\right)  $ and the disorder potential $V\left(
\theta\right)  $ break $SO\left(  2\right)  $ symmetry. Although $J_{z}$ is no
longer conserved, eigenstate of the system $\left\langle \theta|\varphi
_{n}\right\rangle =\exp[i\phi_{n}\left(  \theta\right)  ]/\sqrt{2\pi}$ with
energy $n\cdot\Delta\epsilon+\epsilon_{0}$ has a well defined winding number
\begin{equation}
w_{n}=\int_{0}^{2\pi}\left\langle \varphi_{n}|\theta\right\rangle \left\langle
\theta|-i\partial_{\theta}|\varphi_{n}\right\rangle =n,
\end{equation}
because its phase term $\phi_{n}\left(  \theta\right)  =n\theta+\delta\left(
\theta\right)  $ must satisfy $\phi_{n}\left(  \theta+2\pi\right)  =\phi
_{n}\left(  \theta\right)  +2\pi n$ ($n\in\mathbb{Z}$) to match the periodical
boundary condition. The phase-shift term $\delta\left(  \theta\right)
=[\epsilon_{0}\theta+\int_{0}^{\theta}V\left(  \theta^{\prime}\right)
\mathrm{d}\theta^{\prime}]/\Delta\epsilon$ with $\delta\left(  2\pi\right)
=0$ is independent of $n$, which is possible only for a massless 1D system.
The winding number characterize the total number of counterclockwise turns
that the wavefunction makes around the origin. The coupling between
eigenstates $\left\vert \varphi_{n}\right\rangle $ and $\left\vert
\varphi_{n^{\prime}}\right\rangle $ with different \emph{winding numbers} $n$
and $n^{\prime}$\ by circularly polarized photon with angular momentum $\pm1$
is proportional to
\begin{align}
C_{n,n^{\prime}}^{\pm}  &  =\left\langle \varphi_{n}\right\vert \hat{r}_{\pm
}\left\vert \varphi_{n^{\prime}}\right\rangle =\int_{0}^{2\pi}\left\langle
\varphi_{n}|\theta\right\rangle r\left(  \theta\right)  e^{\pm i\theta
}\left\langle \theta|\varphi_{n^{\prime}}\right\rangle \nonumber\\
&  =r_{0}\delta_{n,n^{\prime}\pm1}+r_{0}\delta_{r}\int_{0}^{2\pi}f\left(
\theta\right)  e^{i\left(  -n+n^{\prime}\pm1\right)  \theta}\mathrm{d}%
\theta/2\pi\text{.}%
\end{align}
The second term of $C_{n,n^{\prime}}^{\pm}$ vanishes when $n=n^{\prime}\pm1$
because $\int_{0}^{2\pi}f\left(  \theta\right)  \mathrm{d}\theta/2\pi=0$. It
can be neglected when $n\neq n^{\prime}\pm1$ because $\int_{0}^{2\pi}f\left(
\theta\right)  \exp[i\left(  -n+n^{\prime}\pm1\right)  \theta]\mathrm{d}%
\theta/2\pi=f_{n^{\prime}\pm1-n}$ is the $\left(  n^{\prime}\pm1-n\right)  $th
Fourier coefficient of $f\left(  \theta\right)  $, and $\left\vert
f_{n^{\prime}\pm1-n}\right\vert \ll1$ for a random fluctuation\cite{Noise}.
Therefore the selection rules for circularly polarized photons in this system
respect winding number conservation $n=n^{\prime}\pm1$. On the other hand, the
edge states in a TIQD are naturally quasi 1D states (see inset in Fig.
\ref{fig2}) with linear dispersions regardless of the TIQD's shape due to
their massless feature, which also originates from the topological property of
the TIQD\cite{Hasan,XLQi}. The winding number for each of the edge states in a
specific basis, e.g., $\{\left\vert e,\uparrow\downarrow\right\rangle ,$
$\left\vert hh,\uparrow\downarrow\right\rangle \}$, is well defined. Therefore
the selection rules between edge states in TIQDs respect the conservation of
the topological winding number. The zero FSS together with the winding number
conservation during the cascade process in TIQDs results in the final
two-photon state, whose entanglement is robust against disorders, morphology
fluctuations and asymmetry of the TIQD.

\textit{Conclusion.--- }In this Letter, we proposed a realizable scheme to
produce photon pairs at terahertz regime with robust entanglement in generic
topological insulator quantum dots (TIQDs). We prove one-dimensional massless
excitons composed of edge states with linear dispersions in TIQDs are free
from the fine structure splitting (FSS) due to the time-reversal symmetry. The
entanglement of the emitted photon pairs from the cascade process is robust
against disorders and morphology fluctuations in TIQDs, because of the absence
of FSS and the selection rules between edge states respecting the winding
number conservation. Both merits root in the topology of the proposed TIQD.
These two findings have \textit{not} been discussed in previous literatures as
far as we know. Note that our results are valid near the Dirac point where the
dispersions of edge states are linear. This discovery paves a new way towards
the future applications and integrations of entangled photon pair emitters,
and possible novel optical devices based on winding number conservation, by
utilizing the massless edge states of topological materials. Our scheme, which
can be implemented using current experimental techniques, is free from the
requirements to fabricate quantum dots with high symmetry and sophisticated
regulations to eliminate FSS.

\begin{acknowledgments}
K.C. is supported by NSFC Grants No.11434010 and the Grant Nos. 2015CB921503,
and 2011CB922204-2 from the MOST of China. C.-P. S. acknowledges support from
the NSFC Grant Nos. 11421063 and 11534002, and the National 973 program (Grant
Nos. 2012CB922104 and 2014CB921403).
\end{acknowledgments}


\begin{thebibliography}{99}
\bibitem {hh}The Four-band Hamiltonian describes the low energy electron
dynamics, which is derived from the full multi-band Hamiltonian with strong
spin-orbit coupling and band mixing effects included. The inversion of the
band gap, i.e., $M>0$ to $M<0$ transition, is the major result of the strong
spin-orbit coupling. $hh$ indicate not exactly but largely heavy hole bands,
as other hole bands are absorbed into the heavy hole bands during the
Hamiltonian reduction.
\bibitem {Noise}It is reasonable to model the fluctuation of a QD's boundary
$f\left(  \theta\right)  $ as the white noise. Because when someone tries to
obtain a QD in circular shape by etching, the boundary of the QD is formed by
atomic dissociation process which is mesoscopically regulated and
microscopically random. Norms of its Fourier coefficients $\left\vert
f_{m}\right\vert \sim\omega_{c}^{-1/2}$, where $\omega_{c}\sim2\pi r_{0}%
/a_{0}$ is the cut-off frequency of the fluctuation, with $r_{0}$ the radius
of the QD and $a_{0}$ the atomic spacing. For realistic QDs, $2\pi r_{0}%
/a_{0}>10^{3}$, $\left\vert f_{m}\right\vert ^{2}<10^{-3}$, and $\left\vert
c_{1}/c_{0}\right\vert ^{2}\sim\delta_{r}^{2}\left\vert f_{m}\right\vert
^{2}<10^{-4}$.
\end{thebibliography}
\end{document}

\title{Supplementary Material: Entangled Terahertz Photon Emission from Topological
Insulator Quantum Dots}
\author{Li-kun Shi}
\affiliation{Beijing Computational Science Research Center, Beijing 100193, China}
\author{Kai Chang}
\affiliation{SKLSM, Institute of Semiconductors, Chinese Academy of Sciences, P.O. Box 912,
Beijing 100083, China}
\author{Chang-Pu Sun}
\affiliation{Beijing Computational Science Research Center, Beijing 100193, China}
\maketitle

\textbf{I. Eigenstates in a Circular HgTe Quantum Dot} \medskip

The low-energy electron states in the HgTe QD are described by a four-band
effective Hamiltonian in the basis $\left\vert e,\uparrow\right\rangle $,
$\left\vert hh,\uparrow\right\rangle $, $\left\vert e,\downarrow\right\rangle
$, and $\left\vert hh,\downarrow\right\rangle $:%
\begin{equation}
H_{4\times4}=\left[
\begin{array}
[c]{cc}%
H_{\uparrow}(\hat{\mathbf{k}}) & 0\\
0 & H_{\downarrow}(\hat{\mathbf{k}})
\end{array}
\right]  +V\left(  r\right)  , \tag{1-1}%
\end{equation}
where $\hat{\mathbf{k}}=(\hat{k}_{x},\hat{k}_{y})$ is the in-plane wavevector,
$H_{\uparrow}(\hat{\mathbf{k}})=H_{\downarrow}^{\ast}(-\hat{\mathbf{k}%
})=\epsilon(\hat{k})I_{2\times2}+d_{i}(\hat{\mathbf{k}})\sigma^{i}$, and
$\sigma^{i}\left(  i=1,2,3\right)  $ are the Pauli matrices; $\epsilon(\hat
{k})=C-D\hat{k}^{2}$, $d_{1,2}(\hat{\mathbf{k}})=A\hat{k}_{x,y} $, and
$d_{3}(\hat{k})=M-B\hat{k}^{2}$. The hard-wall confining potential $V\left(
r\right)  $ is employed to describe potential difference between the vacuum
and the TIQD with $R$ as its radius. The four-band Hamiltonian is derived from
the full multi-band Hamiltonian with strong spin-orbit coupling and band
mixing effects included. The total angular momentum $\hat{J}_{z,\uparrow
\downarrow}=\hat{L}_{z}+\hat{S}_{z,\uparrow\downarrow}$, where $\hat{L}_{z}$
is the orbit azimuthal angular momentum, and $\hat{S}_{z,\uparrow\downarrow}$
is the spin projection onto the $z$ direction, which is $\pm1/2$ for
$\left\vert e,\uparrow\downarrow\right\rangle $ and $\pm3/2$ for $\left\vert
hh,\uparrow\downarrow\right\rangle $. One can verify $[H_{\uparrow\downarrow
},\hat{J}_{z,\uparrow\downarrow}]=0$.

The eigenstate of $\hat{J}_{z}$ in the basis $\{\left\vert e,\uparrow
\right\rangle ,\left\vert hh,\uparrow\right\rangle ,\left\vert e,\downarrow
\right\rangle ,\left\vert hh,\downarrow\right\rangle \}$ satisfying $\hat
{J}_{z,,\uparrow\downarrow}\left\vert \psi_{j,\uparrow\downarrow}\right\rangle
=j\left\vert \psi_{j,\uparrow\downarrow}\right\rangle $ has the form
\begin{equation}
\psi_{j,\uparrow\downarrow}(r,\phi)=\{c_{1,\uparrow\downarrow}J_{j-1/2}\left(
kr\right)  \exp[i\left(  j\mp1/2\right)  \phi],c_{2,\uparrow\downarrow
}J_{j-3/2}\left(  kr\right)  \exp[i\left(  j\mp3/2\right)  \phi]\}^{T}
\tag{1-2}%
\end{equation}
in the polar coordinate. For bulk states with no boundary constraints on $k$,
the eigenequation is $\hat{H}_{\uparrow\downarrow}\left\vert \psi
_{j,\uparrow\downarrow}\right\rangle =\epsilon\left\vert \psi_{j,\uparrow
\downarrow}\right\rangle $, which leads to $\mathcal{M}_{1,\uparrow\downarrow
}[c_{1,\uparrow\downarrow},c_{2,\uparrow\downarrow}]^{T}=0$ and the relation
between $k$ and $\epsilon$:%
\begin{equation}
\det\mathcal{M}_{1,\uparrow\downarrow}=0, \tag{1-3}%
\end{equation}
where
\begin{equation}
\mathcal{M}_{1,\uparrow\downarrow}\mathcal{=}\left[
\begin{array}
[c]{cc}%
M-B_{+}k^{2}-\epsilon & \mp iAk\\
\pm iAk & -M+B_{-}k^{2}-\epsilon
\end{array}
\right]  . \tag{1-4}%
\end{equation}

As the bulk spectrum has two branches, for a given value of $\epsilon$ there
are two non-trivial solutions $k_{i}$ ($i=1,2$) for the momentum. For
eigenstates in the QD, we are able to satisfy their vanishing boundary
condition $\psi_{j,\uparrow\downarrow}\left(  R\right)  =0$ by combining two
linearly independent degenerate solutions in the bulk $\psi_{j,\uparrow
\downarrow}\left(  r\right)  =\sum_{i=1}^{2}[c_{1i,\uparrow\downarrow}%
J_{j\mp1/2}\left(  k_{i}r\right)  ,c_{2i,\uparrow\downarrow}J_{j\mp3/2}\left(
k_{i}r\right)  ]^{T}$ and requiring the coefficients to satisfy $\mathcal{M}%
_{1,\uparrow\downarrow}[c_{1i,\uparrow\downarrow},c_{2i,\uparrow\downarrow
}]^{T}=0$. For convenience we define $\gamma_{2i,\uparrow\downarrow}$ as%
\begin{equation}
\gamma_{2i,\uparrow\downarrow}=c_{2i,\uparrow\downarrow}/c_{1i,\uparrow
\downarrow}=\pm iAk_{i}/(-M+B_{-}k_{i}^{2}-\epsilon). \tag{1-5}%
\end{equation}
Therefore
\begin{equation}
\psi_{j,\uparrow\downarrow}\left(  r\right)  =\sum_{i=1}^{2}c_{1i,\uparrow
\downarrow}[J_{j\mp1/2}\left(  k_{i}r\right)  ,\gamma_{2i,\uparrow\downarrow
}J_{j\mp3/2}\left(  k_{i}r\right)  ]^{T}, \tag{1-6}%
\end{equation}
then $\psi_{j,\uparrow\downarrow}\left(  R\right)  =0$ leads to $\mathcal{M}%
_{2,\uparrow\downarrow}[c_{11,\uparrow\downarrow},c_{12,\uparrow\downarrow
}]^{T}=0$ and the equation to fix the eigenenergy:%
\begin{equation}
\det\mathcal{M}_{2,\uparrow\downarrow}=0\text{,} \tag{1-7}%
\end{equation}
where%
\begin{equation}
\mathcal{M}_{2}=\left[
\begin{array}
[c]{cc}%
J_{j\mp1/2}\left(  k_{1}R\right)  & J_{j\mp1/2}\left(  k_{2}R\right) \\
\gamma_{21,\uparrow\downarrow}J_{j\mp3/2}\left(  k_{1}R\right)  &
\gamma_{22,\uparrow\downarrow}J_{j\mp3/2}\left(  k_{2}R\right)
\end{array}
\right]  . \tag{1-8}%
\end{equation}

\bigskip

\textbf{II. Emissions from the Superconducting }$p$\textbf{-}$n$%
\textbf{\ Junction} \medskip

The weak coupling between the QD and the SC leads induce a pair potential
$\Delta_{c,v}=\exp[i\phi_{\mathrm{s,d}}]\Gamma_{\mathrm{s,d}}/2$ for the
conduction (valance) level, i.e., the proximity effect. $\phi_{\mathrm{s,d}}$
is the phase of the corresponding $\Delta_{\mathrm{s},\mathrm{d}}$, and
$\Gamma_{\mathrm{s,d}}$ characterize the broadening of the conduction
(valance) level proportional to square of the tunneling amplitude (assuming
$\Gamma_{\mathrm{s,d}}\ll\left\vert \Delta_{\mathrm{s},\mathrm{d}}\right\vert
$, i.e., the coupling between the QD and SC leads is weak). The effective
low-energy Hamiltonian for the conduction ($\alpha=c$) or valance ($\alpha=v$)
level is%
\begin{align}
\tilde{H}_{\alpha}  &  =\tilde{\epsilon}_{\alpha}\sum_{\sigma=\uparrow
\downarrow}a_{\alpha,\sigma}^{\dag}a_{\alpha,\sigma}+Un_{\alpha,\uparrow
}n_{\alpha,\downarrow}\nonumber\\
&  +\Delta_{\alpha}a_{\alpha,\uparrow}^{\dag}a_{\alpha,\downarrow}^{\dag
}+\Delta_{\alpha}^{\ast}a_{\alpha,\downarrow}a_{\alpha,\uparrow}, \tag{2-1}%
\end{align}
where $\tilde{\epsilon}_{c,v}=\epsilon_{c,v}-\mu_{\mathrm{s,d}}$ counts from
$\mu_{\mathrm{s},\mathrm{d}}$. By diagonalizing $\tilde{H}_{\alpha}$, beside
two degenerate states $a_{\alpha,\uparrow\downarrow}^{\dag}\left\vert
0\right\rangle _{\alpha}$ with energy $\tilde{\epsilon}_{\alpha}$, there are
excited and ground states being linear superposition of empty state
$\left\vert 0\right\rangle _{\alpha}$ and fully occupied two-particle state
$\left\vert 2\right\rangle _{\alpha}$, i.e.,
\begin{equation}
\left\vert \pm\right\rangle _{\alpha}=-\exp[-i\phi_{\alpha}]\left\vert
u_{\alpha,\pm}\right\vert \left\vert 0\right\rangle _{\alpha}+\left\vert
u_{\alpha,\mp}\right\vert \left\vert 2\right\rangle _{\alpha}, \tag{2-2}%
\end{equation}
with
\begin{equation}
\tilde{\epsilon}_{\alpha,\pm}=\varepsilon_{\alpha}\pm(\varepsilon_{\alpha}%
^{2}+\left\vert \Delta_{\alpha}\right\vert ^{2})^{1/2}, \tag{2-3}%
\end{equation}
where $\varepsilon_{\alpha}=\tilde{\epsilon}_{\alpha}+U/2$ and $\left\vert
u_{\alpha,\pm}\right\vert =(1/\sqrt{2})\times\lbrack1\pm\varepsilon_{\alpha
}/(\varepsilon_{\alpha}^{2}+\left\vert \Delta_{\alpha}\right\vert ^{2}%
)^{1/2}]^{1/2}$. \begin{figure}[tbh]
\includegraphics[width=0.86\columnwidth]{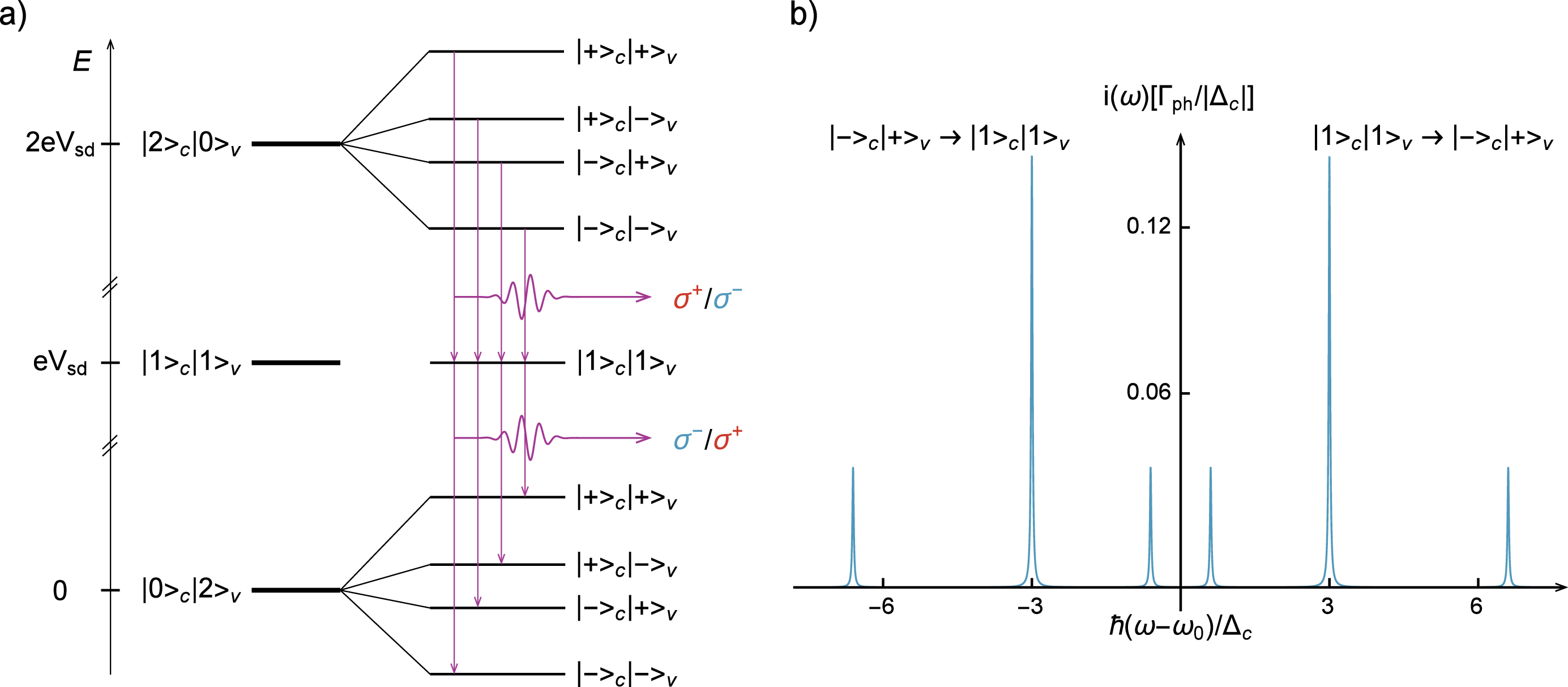}\caption{(a) Sketch of the
energy level splittings for two-particle states in the HgTe QD weakly coupled
to SC electrodes. The $p$-$n$ junction is biased with a voltage
$V_{\mathrm{s,d}}$. The cascade produces two photons with opposite circular
polarization $\sigma^{\pm}$. (b) Emission intensity of time- and
polarization-correlated photons emitted from the biexciton cascade. The
central frequency $\hbar\omega_{0}\simeq\mathrm{e}V_{\mathrm{sd}}$. The
dominant cascade process is $\left\vert -\right\rangle _{c}\left\vert
+\right\rangle _{v}\rightarrow\left\vert 1\right\rangle _{c}\left\vert
1\right\rangle _{v}\rightarrow\left\vert -\right\rangle _{c}\left\vert
+\right\rangle _{v}$. Parameters: $\tilde{\epsilon}_{c}=\tilde{\epsilon}%
_{v}=0$, $U=3$, $\Gamma_{\mathrm{ph}}=0.02$ (in units of $\Delta_{c}%
=\Delta_{v}$).}%
\label{fig1s}%
\end{figure}

The emission intensity computed versus photon frequency $\omega$ is shown in
Fig. \ref{fig1s} (b). The emission intensity is $i\left(  \omega_{q_{z}%
}\right)  =\sum_{ab,p}w_{a\rightarrow b}^{p}\left(  \omega_{q_{z}}\right)
\rho_{a}$ with the transition rate $w_{a\rightarrow b}^{p}\left(
\omega_{q_{z}}\right)  =(2\pi/\hbar)\sum_{q_{z}}\left\vert \left\langle
b;q_{z},p\left\vert H_{\mathrm{int}}\right\vert a;0\right\rangle \right\vert
^{2}\delta\lbrack\tilde{\epsilon}_{ab}-\hbar\tilde{\omega}_{q_{z}}%
+i\Gamma_{\mathrm{ph}}]$. We will explain below the symbols in $i\left(  \omega_{q_{z}%
}\right)  $ and $w_{a\rightarrow b}^{p}\left(  \omega_{q_{z}}\right)  $:

$a$, $b$ are one of six possible states in the QD \{$\left\vert +\right\rangle
_{c}\left\vert +\right\rangle _{v}$, $\left\vert +\right\rangle _{c}\left\vert
-\right\rangle _{v}$, $\left\vert -\right\rangle _{c}\left\vert +\right\rangle
_{v}$, $\left\vert -\right\rangle _{c}\left\vert -\right\rangle _{v}$,
$\left\vert \uparrow\right\rangle _{c}\left\vert \downarrow\right\rangle _{v}%
$, $\left\vert \downarrow\right\rangle _{c}\left\vert \uparrow\right\rangle
_{v}$\} [see Fig. \ref{fig1s} (a)]. These six states are direct products of
two eigenstates (one from the conduction level and another one from the
valance level) and can always be expanded by in the basis \{$\left\vert
2\right\rangle _{c}\left\vert 2\right\rangle _{v}$,$\left\vert 0\right\rangle
_{c}\left\vert 0\right\rangle _{v}$,$\left\vert 2\right\rangle _{c}\left\vert
0\right\rangle _{v}$,$\left\vert 0\right\rangle _{c}\left\vert 2\right\rangle
_{v}$,$\left\vert \uparrow\right\rangle _{c}\left\vert \downarrow\right\rangle
_{v}$,$\left\vert \downarrow\right\rangle _{c}\left\vert \uparrow\right\rangle
_{v}$\}. In our work we only use electron notations and do not introduce hole
operators. For example, $\left\vert 2\right\rangle _{c}\left\vert
0\right\rangle _{v}$ means there are two electrons in the conduction level and
no electron in the valance level (biexciton); $\left\vert 1\right\rangle
_{c}\left\vert 1\right\rangle _{v}$ means there is one electron in the
conduction level and one electron in the valance level (exciton); $\left\vert
0\right\rangle _{c}\left\vert 2\right\rangle _{v}$ means there is no electron
in the conduction level and two electrons in the valance level (no exciton).

$p=\sigma^{\pm}$ denotes the polarization; $\tilde{\epsilon}_{ab}%
=\tilde{\epsilon}_{a}-\tilde{\epsilon}_{b}$ is the energy difference between
states $a$ and $b$ (Note that $\tilde{\epsilon}_{a}$ or $\tilde{\epsilon}_{b}$
is the energy for \textit{two-particle} state); $\tilde{\omega}_{q_{z}}%
=\omega_{q_{z}}-\omega_{0}$ is the frequency shift caused by proximity effect;
and $\Gamma_{\mathrm{ph}}=2\pi\nu_{\mathrm{ph}}\left\vert g\right\vert ^{2}$
is the broadening of the emission peaks with $\nu_{\mathrm{ph}}$ the photon
DOS assumed to be independent of energy and polarization. The probabilities
$\rho_{a}$ are determined by the stationary solution of the master equation
$\dot{\rho}_{a}=\sum_{b,p}[W_{b\rightarrow a}^{p}\rho_{b}-W_{a\rightarrow
b}^{p}\rho_{a}]$, $W_{a\rightarrow b}^{p}=\int\mathrm{d}\omega^{\prime
}w_{a\rightarrow b}^{p}\left(  \omega^{\prime}\right)  $.

At the first glance, it is weird to see the initial state of the QD is the
same with the final state after biexciton cascade. In fact, the state
$\left\vert -\right\rangle _{c}\left\vert +\right\rangle _{v}$ is the
superposition of two parts: $\left\vert 2\right\rangle _{c}\left\vert
0\right\rangle _{v}$ with two electrons in the conduction level and
$\left\vert 0\right\rangle _{c}\left\vert 2\right\rangle _{v}$ with two
electrons in the valance level. Therefore the initial state is actually the
$\left\vert 2\right\rangle _{c}\left\vert 0\right\rangle _{v}$ part of the
$\left\vert -\right\rangle _{c}\left\vert +\right\rangle _{v}$ and the final
state is the $\left\vert 0\right\rangle _{c}\left\vert 2\right\rangle _{v}$
part of the $\left\vert -\right\rangle _{c}\left\vert +\right\rangle _{v}$.
However, with induced pair potential $\Delta_{c,v}$, $\left\vert
2\right\rangle _{c}\left\vert 0\right\rangle _{v}$ and $\left\vert
0\right\rangle _{c}\left\vert 2\right\rangle _{v}$ are no longer two-particle
eigenstates. One has to calculate the process like $\left\vert -\right\rangle
_{c}\left\vert +\right\rangle _{v}\rightarrow\left\vert 1\right\rangle
_{c}\left\vert 1\right\rangle _{v}\rightarrow\left\vert -\right\rangle
_{c}\left\vert +\right\rangle _{v}$ instead of $\left\vert 2\right\rangle
_{c}\left\vert 0\right\rangle _{v}\rightarrow\left\vert 1\right\rangle
_{c}\left\vert 1\right\rangle _{v}\rightarrow\left\vert 0\right\rangle
_{c}\left\vert 2\right\rangle _{v}$, while the frequencies of the emitter
photons are shifted from the central frequency $\hbar\omega_{0}=(2\epsilon
_{c}+U)-(\epsilon_{c}+\epsilon_{v}+U)=(\epsilon_{c}+\epsilon_{v}%
+U)-(2\epsilon_{v}+U)=\epsilon_{c}-\epsilon_{v}=\mathrm{e}V_{\mathrm{sd}}$,
where $2\epsilon_{c}+U$ is the energy for $\left\vert 2\right\rangle
_{c}\left\vert 0\right\rangle _{v}$ state; $2\epsilon_{v}+U$ is the energy for
$\left\vert 2\right\rangle _{c}\left\vert 0\right\rangle _{v}$ state;
$(\epsilon_{c}+\epsilon_{v}+U)$ is the energy for $\left\vert 1\right\rangle
_{c}\left\vert 1\right\rangle _{v}$ state.

\bigskip

\textbf{III. Coulomb Interactions between Two Particle States} \medskip

The two-particle Hamiltonian for the conduction and valance level is
\begin{equation}
H_{cv}=\sum_{\alpha=c,v}\sum_{\sigma=\uparrow\downarrow}\epsilon_{\alpha
}a_{\alpha,\sigma}^{\dag}a_{\alpha,\sigma}+\hat{V}, \tag{3-1}%
\end{equation}
where $\hat{V}$ is the Coulomb interaction between the two electrons. Any
exciton state must be the linear combinations of the two-particle basis
$\{\left\vert \uparrow\right\rangle _{c}\left\vert \downarrow\right\rangle
_{v},\left\vert \downarrow\right\rangle _{c}\left\vert \uparrow\right\rangle
_{v},\left\vert \uparrow\right\rangle _{c}\left\vert \uparrow\right\rangle
_{v},\left\vert \downarrow\right\rangle _{c}\left\vert \downarrow\right\rangle
_{v}\}$, in which one electron is in the conduction level and another in the
valance level. These two-particle states are products of edge states which
satisfy: 1) $\left\vert \downarrow\right\rangle _{\alpha}=\left\vert
\Theta\uparrow\right\rangle _{\alpha}$ ($\alpha=c,v$) due to the TRS; 2) an
additional symmetry $\left\vert \uparrow\downarrow\right\rangle _{c}%
=i\Theta\left\vert \uparrow\downarrow\right\rangle _{v}$ as the conduction and
valance level are evenly away from and near the Dirac point. These two
relations hold, not only for HgTe QDs, but also for generic TIQDs that are
composed of other materials and can have irregular shapes.

In the two-particle basis $\{\left\vert \uparrow\right\rangle _{c}\left\vert
\downarrow\right\rangle _{v},\left\vert \downarrow\right\rangle _{c}\left\vert
\uparrow\right\rangle _{v},\left\vert \uparrow\right\rangle _{c}\left\vert
\uparrow\right\rangle _{v},\left\vert \downarrow\right\rangle _{c}\left\vert
\downarrow\right\rangle _{v}\}$,%
\begin{equation}
H_{cv}=\left[
\begin{array}
[c]{cccc}%
\epsilon_{c}+\epsilon_{v}+U & 0 & 0 & 0\\
0 & \epsilon_{c}+\epsilon_{v}+U & 0 & 0\\
0 & 0 & \epsilon_{c}+\epsilon_{v}+U^{\prime} & \left\langle \uparrow
\right\vert _{c}\left\langle \uparrow\right\vert _{v}\hat{V}\left\vert
\downarrow\right\rangle _{c}\left\vert \downarrow\right\rangle _{v}\\
0 & 0 & \left\langle \downarrow\right\vert _{c}\left\langle \downarrow
\right\vert _{v}\hat{V}\left\vert \uparrow\right\rangle _{c}\left\vert
\uparrow\right\rangle _{v} & \epsilon_{c}+\epsilon_{v}+U^{\prime}%
\end{array}
\right]  , \tag{3-2}%
\end{equation}
where%
\begin{align}
U  &  =\left\langle \uparrow\right\vert _{c}\left\langle \downarrow\right\vert
_{v}\hat{V}\left\vert \uparrow\right\rangle _{c}\left\vert \downarrow
\right\rangle _{v}=\left\langle \Theta\downarrow\right\vert _{c}\left\langle
-\Theta\uparrow\right\vert _{v}\hat{V}\left\vert \Theta\downarrow\right\rangle
_{c}\left\vert -\Theta\uparrow\right\rangle _{v}=\left\langle \downarrow
\right\vert _{c}\left\langle \uparrow\right\vert _{v}\hat{V}\left\vert
\downarrow\right\rangle _{c}\left\vert \uparrow\right\rangle _{v},\tag{3-3}\\
U^{\prime}  &  =\left\langle \uparrow\right\vert _{c}\left\langle
\uparrow\right\vert _{v}\hat{V}\left\vert \uparrow\right\rangle _{c}\left\vert
\uparrow\right\rangle _{v}=\left\langle \Theta\downarrow\right\vert
_{c}\left\langle -\Theta\downarrow\right\vert _{v}\hat{V}\left\vert
\Theta\downarrow\right\rangle _{c}\left\vert -\Theta\downarrow\right\rangle
_{v}=\left\langle \uparrow\right\vert _{c}\left\langle \uparrow\right\vert
_{v}\hat{V}\left\vert \downarrow\right\rangle _{c}\left\vert \downarrow
\right\rangle _{v}.\nonumber
\end{align}

All but two off-diagonal terms in $H_{cv}$ are zero. For example,%
\begin{equation}
\left\langle \uparrow\right\vert _{c}\left\langle \downarrow\right\vert
_{v}H_{cv}\left\vert \downarrow\right\rangle _{c}\left\vert \uparrow
\right\rangle _{v}=\left\langle \uparrow\right\vert _{c}\left\langle
\downarrow\right\vert _{v}\hat{V}\left\vert \downarrow\right\rangle
_{c}\left\vert \uparrow\right\rangle _{v}=I_{1}-I_{2}=0, \tag{3-4}%
\end{equation}
in which $\left\vert \uparrow\right\rangle _{c}\left\vert \downarrow
\right\rangle _{v}=[\left\vert \uparrow_{1}\right\rangle _{c}\left\vert
\downarrow_{2}\right\rangle _{v}-\left\vert \uparrow_{2}\right\rangle
_{c}\left\vert \downarrow_{1}\right\rangle _{v}]/\sqrt{2}$, $\left\vert
\downarrow\right\rangle _{c}\left\vert \uparrow\right\rangle _{v}=[\left\vert
\downarrow_{1}\right\rangle _{c}\left\vert \uparrow_{2}\right\rangle
_{v}-\left\vert \downarrow_{2}\right\rangle _{c}\left\vert \uparrow
_{1}\right\rangle _{v}]/\sqrt{2}$ where we assign $1$ and $2$ explicitly to
label two identical electrons;%
\begin{align}
I_{1}  &  =\left\langle \uparrow_{1}\right\vert _{c}\left\langle
\downarrow_{2}\right\vert _{v}V_{12}\left\vert \downarrow_{1}\right\rangle
_{c}\left\vert \uparrow_{2}\right\rangle _{v}=\left\langle \uparrow
_{1}\right\vert _{c}\left\langle \downarrow_{2}\right\vert _{v}V_{12}%
\left\vert \Theta\uparrow_{1}\right\rangle _{c}\left\vert \uparrow
_{2}\right\rangle _{v}\nonumber\\
&  =\left\langle \Theta^{2}\uparrow_{1}\right\vert _{c}\left\langle
\downarrow_{2}\right\vert _{v}V_{12}\left\vert \Theta\uparrow_{1}\right\rangle
_{c}\left\vert \uparrow_{2}\right\rangle _{v}=-I_{1}=0, \tag{3-5}%
\end{align}%
\begin{align}
I_{2}  &  =\left\langle \uparrow_{1}\right\vert _{c}\left\langle
\downarrow_{2}\right\vert _{v}V_{12}\left\vert \downarrow_{2}\right\rangle
_{c}\left\vert \uparrow_{1}\right\rangle _{v}=\left\langle i\Theta\uparrow
_{1}\right\vert _{v}\left\langle \downarrow_{2}\right\vert _{v}V_{12}%
\left\vert i\Theta\downarrow_{2}\right\rangle _{v}\left\vert \uparrow
_{1}\right\rangle _{v}\nonumber\\
&  =\left\langle \Theta\uparrow_{1}\right\vert _{v}\left\langle \downarrow
_{2}\right\vert _{v}V_{12}\left\vert \Theta\downarrow_{2}\right\rangle
_{v}\left\vert \Theta^{2}\uparrow_{1}\right\rangle _{v}=-I_{2}=0. \tag{3-6}%
\end{align}
Note that $V_{12}=V_{21}$ is the Coulomb interaction between the electron 1
and 2 and $V_{12}$ commutes with the time-reversal operator $\Theta^{2}=-1$
for fermions. Other zero off-diagonal terms can be obtained by similar analysis.

Among the four two-particle states $\{\left\vert \uparrow\right\rangle
_{c}\left\vert \downarrow\right\rangle _{v},\left\vert \downarrow\right\rangle
_{c}\left\vert \uparrow\right\rangle _{v},\left\vert \uparrow\right\rangle
_{c}\left\vert \uparrow\right\rangle _{v},\left\vert \downarrow\right\rangle
_{c}\left\vert \downarrow\right\rangle _{v}\}$, $\left\vert \uparrow
\right\rangle _{c}\left\vert \uparrow\right\rangle _{v}$ and $\left\vert
\downarrow\right\rangle _{c}\left\vert \downarrow\right\rangle _{v}$ are dark
states. The optical transitions within $\left\vert \uparrow\right\rangle
_{c}\left\vert \uparrow\right\rangle _{v}$ and $\left\vert \downarrow
\right\rangle _{c}\left\vert \downarrow\right\rangle _{v}$ are not allowed
because the spin-up (down) states are fully occupied in $\left\vert
\uparrow\right\rangle _{c}\left\vert \uparrow\right\rangle _{v}$ ($\left\vert
\downarrow\right\rangle _{c}\left\vert \downarrow\right\rangle _{v}$) state,
while transitions between opposite spin states are forbidden.

These two dark states are irrelevant to our cascade process, and do not couple
with other two bright states. Therefore, the bright exciton states must be
linear combinations of two-particle bright states $\left\vert \uparrow
\right\rangle _{c}\left\vert \downarrow\right\rangle _{v}$ and $\left\vert
\downarrow\right\rangle _{c}\left\vert \uparrow\right\rangle _{v}$.

In the two-particle basis $\{\left\vert \uparrow\right\rangle _{c}\left\vert
\downarrow\right\rangle _{v},\left\vert \downarrow\right\rangle _{c}\left\vert
\uparrow\right\rangle _{v}\}$, as we analyzed above,%
\begin{equation}
H_{cv}=\left[
\begin{array}
[c]{cc}%
\epsilon_{c}+\epsilon_{v}+U & 0\\
0 & \epsilon_{c}+\epsilon_{v}+U
\end{array}
\right]  , \tag{3-7}%
\end{equation}
therefore $\left\vert \uparrow\right\rangle _{c}\left\vert \downarrow
\right\rangle _{v}$ and $\left\vert \downarrow\right\rangle _{c}\left\vert
\uparrow\right\rangle _{v}$ are degenerate bright exciton states with the same
energy $\epsilon_{c}+\epsilon_{v}+U$.

The Coulomb repulsion between two electrons in the same conduction/valance
level is
\begin{equation}
U_{c}=\left\langle \uparrow\right\vert _{c}\left\langle \downarrow\right\vert
_{c}\hat{V}\left\vert \uparrow\right\rangle _{c}\left\vert \downarrow
\right\rangle _{c}=\left\langle i\Theta\uparrow\right\vert _{v}\left\langle
i\Theta\downarrow\right\vert _{v}\hat{V}\left\vert i\Theta\uparrow
\right\rangle _{v}\left\vert i\Theta\downarrow\right\rangle _{v}=U_{v},
\tag{3-8}%
\end{equation}
while the Coulomb repulsion between one electron in the conduction level and
another electron in the valance level is%
\begin{equation}
U=\left\langle \downarrow\right\vert _{c}\left\langle \uparrow\right\vert
_{v}\hat{V}\left\vert \downarrow\right\rangle _{c}\left\vert \uparrow
\right\rangle _{v}=\left\langle i\Theta\downarrow\right\vert _{v}\left\langle
\uparrow\right\vert _{v}\hat{V}\left\vert i\Theta\downarrow\right\rangle
_{v}\left\vert \uparrow\right\rangle _{v}=U_{v}, \tag{3-9}%
\end{equation}
therefore $U_{c,v}=U$.

\bigskip

\textbf{IV. Notes on Finite Element Method Simulations} \medskip

To obtain the edge states in the deformed HgTe QD, we still start from the
four-band effective Hamiltonian in the basis $\left\vert e,\uparrow
\right\rangle $, $\left\vert hh,\uparrow\right\rangle $, $\left\vert
e,\downarrow\right\rangle $, and $\left\vert hh,\downarrow\right\rangle $:%
\begin{equation}
H_{4\times4}=\left[
\begin{array}
[c]{cc}%
H_{\uparrow}(\hat{\mathbf{k}}) & 0\\
0 & H_{\downarrow}(\hat{\mathbf{k}})
\end{array}
\right]  +V\left(  r\right)  , \tag{4-1}%
\end{equation}
but here $V\left(  r\right)  $ is the hard-wall potential with its boundary
described by $r\left(  \theta\right)  =r_{0}[1+\delta_{r}f\left(
\theta\right)  ]$ in the polar coordinate system, with the mean radius
$r_{0}=\int_{0}^{2\pi}r\left(  \theta\right)  \mathrm{d}\theta/2\pi$ and the
random function $f\left(  \theta\right)  \in\left[  -1,1\right]  $ satisfies
$f\left(  \theta\right)  =f\left(  \theta+2\pi\right)  $. $\delta_{r}$ is a
dimensionless number characterizing the maximum fluctuation percentage. Each
random function for each of the QDs is generated independently as%
\begin{equation}
f\left(  \theta\right)  =(1/\mathcal{N})\sum_{n=1}^{N}c_{i}\sin(n\theta
+\varphi_{i})/\sqrt{\pi}, \tag{4-2}%
\end{equation}
where $N$ is generated randomly in the interval $[10,100]$; each $c_{i}%
$($\varphi_{i})$ is generated randomly and independently in the interval
$\left[  0,2\pi\right]  $; $\mathcal{N=}\sum_{n=1}^{N}c_{i}^{2}$. This
generating scheme makes $f\left(  \theta\right)  $ to be a white noise. It is
reasonable to model the fluctuation of a QD's boundary $f\left(
\theta\right)  $ as the white noise. Because the boundary of a QD formed by
etching is caused by atomic dissociation process which is mesoscopically
regulated and microscopically random.

When the Hamiltonian of a deformed QD is generated, we solve the
Schr\"{o}dinger equation for the QD with finite element method, and obtain the
wavefunctions for the edge states$\left\vert \uparrow\downarrow\right\rangle
_{c}$ and $\left\vert \uparrow\downarrow\right\rangle _{v}$. Then we can
calculate $\left\vert c_{1}/c_{0}\right\vert ^{2}$ as%
\begin{equation}
\left\vert \frac{c_{1}}{c_{0}}\right\vert ^{2}=\left\vert \frac{\left\langle
\uparrow\right\vert _{v}\hat{r}_{+}\left\vert \uparrow\right\rangle _{c}%
}{\left\langle \uparrow\right\vert _{v}\hat{r}_{-}\left\vert \uparrow
\right\rangle _{c}}\right\vert ^{2}=\left\vert \frac{\left\langle
\downarrow\right\vert _{v}\hat{r}_{-}\left\vert \downarrow\right\rangle _{c}%
}{\left\langle \downarrow\right\vert _{v}\hat{r}_{+}\left\vert \downarrow
\right\rangle _{c}}\right\vert ^{2}. \tag{4-3}%
\end{equation}